\documentclass{amsart}

 \newtheorem{thm}{Theorem}
 
 \newtheorem{lemma}[thm]{Lemma}
 \newtheorem{prop}[thm]{Proposition}
 \newtheorem{theorem}[thm]{Theorem}
 \theoremstyle{definition}
 \newtheorem{defn}[thm]{Definition}
 \theoremstyle{remark}

 \newcommand{\Real}{\mathbb{R}}

\usepackage{graphicx}

\begin{document}

\title[Fluxon waves in annular Josephson junctions]
{Existence, uniqueness and multiplicity of rotating fluxon waves in annular Josephson junctions}

\author{Guy Katriel}
\address{Einstein Institute of Mathematics, The Hebrew University of Jerusalem, Jerusalem, 91904, Israel}
\email{haggaik@wowmail.com}



\dedicatory{}
\thanks{Partially supported by the Edmund Landau
Center for Research in Mathematical Analysis and Related Areas,
sponsored by the Minerva Foundation (Germany).}

\begin{abstract}
We prove that the equation modelling an annular Josephson junction
has a rotating fluxon wave solution for all values of the
parameters. We also obtain results on uniqueness of the rotating
fluxon wave in some parameter regimes, and on multiplicity of
rotating fluxon waves in other parameter regimes.

\end{abstract}

\maketitle

\section{Introduction}

The equation modelling a long Josephson junction is
\cite{mclaughlin}
\begin{equation}\label{jo}
\phi_{tt}+\alpha\phi_t+\sin(\phi)=\phi_{xx}+\beta\phi_{xxt}+\gamma,
\end{equation}
where $\alpha>0,\beta>0,\gamma>0$ ($\alpha,\beta$ describe
dissipative effects, while $\gamma$ describes the applied bias
current). In the case of annular geometry the periodicity condition
\begin{equation}\label{bnd}
\phi(x+L,t)=\phi(x,t)+2\pi,
\end{equation}
is imposed, where $L$ is the circumference of the junction.

\begin{defn}A {\it{rotating fluxon wave}} is a solution of (\ref{jo}), (\ref{bnd})
of the form
\begin{equation}\label{form}
\phi(x,t)=\theta(kx+\omega t),
\end{equation}
where the function $\theta(z)$ satisfies
\begin{equation}\label{percond}
\theta(z+2\pi)=\theta(z)+2\pi\;\;\; \forall z\in \Real.
\end{equation}
\end{defn}

Clearly the assumption that (\ref{form}) satisfies (\ref{bnd})
implies that
\begin{equation}\label{defk}
k=\frac{2\pi}{L},
\end{equation}
while the frequency $\omega$ is an unknown to be found as part of
the solution (in Proposition \ref{peri} we show that necessarily
$\omega>0$). We note that the time-period $T$ of rotation of the
wave around the annulus is given by $T=\frac{2\pi}{\omega}$.

Substituting (\ref{form}) into (\ref{jo}) we obtain
\begin{equation}
\label{nu} \beta \omega k^2
\theta'''(z)+(k^2-\omega^2)\theta''(z)-\alpha\omega
\theta'(z)-\sin(\theta(z))+\gamma=0,
\end{equation}
so that rotating fluxon waves are given by solutions
$(\omega,\theta(z))$ of (\ref{nu}), with $\theta(z)$ satisfying
(\ref{percond}).

Our aim in this paper is to study the questions of existence and
uniqueness/multiplicity of rotating fluxon waves. We shall obtain
existence of a rotating fluxon wave for all values of the
parameters:
\begin{theorem}
\label{main} For any $L>0$, $\alpha> 0$, $\beta> 0$, $\gamma>0$,
(\ref{jo}), (\ref{bnd}) has a rotating fluxon wave solution
(\ref{form}), with $\omega>0$.
\end{theorem}
To prove Theorem \ref{main} and our other results, we use methods of
nonlinear functional analysis, reducing the problem to a fixed-point
equation depending on a parameter ($\omega$), to be solved by means
of Leray-Schauder theory, coupled with an auxilliary scalar-valued
equation, which we analyze. We have used a similar technique to
prove existence of travelling waves of the {\it{discrete}} damped
and dc-driven sine-Gordon equation \cite{katriel}. The details of
the analysis differ, and in the discrete sine-Gordon case we could
not prove existence for all parameter values (which reflects the
phenomenon of `pinning' in the discrete case) - while here we are
able to do so.

Several researchers have studied long Josephson junctions, and
annular ones in particular, experimentally, \cite{davidson, ust,
ustinov}, numerically \cite{brown, maksimov} and analytically
\cite{derks, hauck,maksimov, mclaughlin}.

Note that in the case $\beta=0$ (known as the damped dc-driven
sine-Gordon equation), (\ref{nu}) reduces to the damped dc-driven
pendulum equation which is amenable to phase-plane analysis and is
thus well understood, and the existence of rotating fluxon waves for
all $\alpha>0, \gamma>0$ is known (see \cite{watanabe}, Section
2.2). The situation for $\beta>0$ is different. Most existing
analytical studies assume that either $\alpha$, $\beta$ and $\gamma$
are small, so that the equation is treated as a (singular)
perturbation of the sine-Gordon equation (the case $\alpha=0$,
$\beta=0$, $\gamma=0$ of (\ref{jo})), or that $\beta$ is small so
that the equation is a perturbation of the damped dc-driven
sine-Gordon equation. An exception is \cite{maksimov}, which studies
travelling fluxon waves of kink type on an infinite line, that is,
solutions of (\ref{jo}) of the form (\ref{form}) with
(\ref{percond}) replaced by $\theta(+\infty)-\theta(-\infty)=2\pi$.
Existence results for such waves are proved by a geometric analysis
of the corresponding three-dimensional phase-space, and a very
intricate structure of solutions is revealed. Here, as we noted, our
approach is functional-analytic rather than geometric, leading to
the general existence result of Theorem \ref{main}.

Our method also allows us to derive some further information about
the set of rotating fluxon waves. In the physical literature, a
common and useful way to describe the behavior of a system is the
`I-V characteristic' - in this case the $\gamma-\omega$ (bias
current-frequency) characteristic:

\begin{defn}
The {\it{bias-frequency characterstic}} is the set $\chi\subset
(0,\infty)\times (0,\infty)$ of all pairs $(\omega,\gamma)$ for
which (\ref{jo}),(\ref{bnd}) has a rotating fluxon wave
(\ref{form}), with frequency $\omega$.
\end{defn}

The set $\chi$ depends, of course, on the parameters
$L,\alpha,\beta$. The following theorem shows that the set $\chi$ is
`large'.

\begin{theorem}
\label{curt} For any $L>0$, $\alpha> 0$, $\beta> 0$, the
bias-frequency characteristic $\chi$ has a subset $\chi_0\subset
\chi$ which has the following properties:

\noindent (i) $\chi_0$ is connected.

\noindent (ii) For any $\omega>0$ there exists $\gamma>0$ with
$(\omega,\gamma)\in \chi_0$.

\noindent (iii) For any $\gamma>0$ there exists $\omega>0$ with
$(\omega,\gamma)\in \chi_0$.

\noindent (iv) $(0,0)\in \overline{\chi_0}$.

\noindent (v) $\chi \setminus\chi_0$ is a bounded set.
\end{theorem}

Note that Theorem \ref{main} follows at once from part (iii) of
Theorem \ref{curt}.

We shall also obtain some bounds on the frequency $\omega$ of
rotating fluxon waves (see section \ref{existence}):
\begin{prop}
\label{peri} For {\bf{any}} rotating fluxon wave of (\ref{jo}),
(\ref{bnd}), we have $\omega>0$ and
\begin{equation}\label{in1}
\frac{\gamma -1}{\alpha}\leq\omega\leq\frac{\gamma}{\alpha}.
\end{equation}
\end{prop}

We note that since we are assured that $\omega>0$, the lower bound
in (\ref{in1}) is nontrivial only when $\gamma>1$.

The above results are completely general in that they hold for all
values of the parameters. We also obtain some further results under
some restrictions on the parameters. While we do not expect our
conditions on the parameters, which are imposed by our methods of
proof, to be sharp, these more restricted results expose some of the
phenomena which can occur and point the way to further investigation
(analytical or numerical) by raising various questions which remain
unresolved.

We do not know whether in general we can take $\chi_0=\chi$ in
Theorem \ref{curt}, or in other words whether the characteristic
$\chi$ is connected. In Section \ref{smoothness} we shall provide a
positive answer to this question, and also obtain further
information about the bias-frequency characteristic $\chi$, for a
certain subset of the space of parameters $L,\alpha,\beta$:

\begin{theorem}\label{connect}
Assume $L>0,\alpha>0,\beta>0$. If
\begin{equation}\label{k1} k=\frac{2\pi}{L}>1
\end{equation}
and
\begin{equation}\label{incon2}
\alpha+\beta k^2\geq \sqrt{2}k
\end{equation}
then the bias-frequency characteristic $\chi$ can be represented
in the form
\begin{equation}\label{rc}
\chi=\{(\omega,\overline{\gamma}(\omega))\;|\; \omega>0\}
\end{equation}
where $\overline{\gamma}:[0,\infty)\rightarrow [0,\infty)$ is
continuous and real-analytic in $(0,\infty)$, and satisfies
\begin{equation}\label{st1}\overline{\gamma}(0)=0,\end{equation}
\begin{equation}\label{st2}\lim_{\omega\rightarrow
+\infty}{\overline{\gamma}}(\omega)=+\infty.\end{equation} In
particular, $\chi$ is connected, and for a given bias $\gamma>0$ the
frequencies of rotating fluxon waves are the solutions  $\omega$ of
the equation
\begin{equation}\label{ceq}
\overline{\gamma}(\omega)=\gamma.
\end{equation}
\end{theorem}

Let us note that in the case $\beta=0$, part of the above theorem is
{\it{not}} true: while $\chi$ is connected and can be parameterized
as in (\ref{rc}) (for any $k>0$, $\alpha>0$; see \cite{watanabe},
sec. 2.2), the function $\overline{\gamma}$ is {\it{not}} analytic
in the case $\beta=0$, in fact it has a cusp at $\omega=k$ (see
\cite{watanabe}, fig. 2), and the value $\omega=k$ separates the
`low-voltage' and `high-voltage' regions. Theorem \ref{connect}
shows that the term $\phi_{xxt}$ in (\ref{jo}) has a smoothing
effect on the curve $\chi$ (at least in the case that
(\ref{k1}),(\ref{incon2}) hold). When we prove Theorem \ref{connect}
we will point out why the assumption $\beta>0$ is crucial, which
leads to this difference between the damped dc-forced sine-Gordon
equation and (\ref{jo}).

A natural question is whether the rotating fluxon wave whose
existence is guaranteed by Theorem \ref{main}, is unique (for given
$L,\alpha,\beta,\gamma$). In Section \ref{uniqueness} we obtain a
uniqueness result in the case that $\gamma$ is sufficiently large.

\begin{theorem}\label{uni1}
Fix any $L>0$, $\alpha>0$, $\beta>0$. Then for sufficiently large
$\gamma>0$, there is a {\bf{unique}} rotating fluxon wave.
\end{theorem}

On the other hand, in Section \ref{multiplicity} we show that
uniqueness does not hold in general, by proving that for certain
parameter ranges there exist at least three rotating fluxon waves
with different frequencies.

\begin{theorem}
\label{multi} Assume (\ref{k1}), and replace (\ref{incon2}) by the
stronger assumption
\begin{equation}\label{lcond2}
\beta k\geq \sqrt{2}.
\end{equation}
Then there exists a value $\gamma_0>0$ such that given any
$0<\epsilon<\frac{1}{2}\gamma_0$, there exists $\alpha_0>0$ so that
for any $\alpha \in (0,\alpha_0)$ and any $\gamma\in
(\epsilon,\gamma_0-\epsilon)$, (\ref{jo}), (\ref{bnd}) has {\bf{at
least}} three rotating fluxon solutions with distinct frequencies.
\end{theorem}

We can expect this multiplicity of rotating fluxon waves to lead to
bistability, jump and hysteresis phenomena as the bias $\gamma$ is
varied. We do not know whether the result of Theorem \ref{multi} is
valid for all values of the parameters $L,\alpha,\beta$, or whether,
on the contrary, there are some values of these parameters for which
there is a {\it{unique}} rotating fluxon wave for {\it{all}}
$\gamma>0$.

We mention here that all our results and proofs in this paper remain
valid, with minor changes, if the nonlinearity $\sin(\phi)$ in
(\ref{jo}) is replaced by $p(\phi)$, where $p$ is an arbitrary $C^1$
$2\pi$-periodic function, satisfying
$\int_0^{2\pi}p(\theta)d\theta=0$.

An important issue which we do not address here is the question of
stability. For example: is it true that for any value of
$\alpha>0,\beta>0,\gamma>0$ there exists at least one rotating
fluxon wave of (\ref{jo}) which is at least locally asymptotically
stable?

\section{Existence of rotating fluxon waves}
\label{existence}

In this section we prove Theorem \ref{curt}, which in particular
implies Theorem \ref{main}. To do so we first recast our problem
into a functional-analytic form, so that we can apply Leray-Schauder
theory.

Multiplying both sides of (\ref{nu}) by $\theta'(z)$ and integrating
over $[0,2\pi]$, taking (\ref{percond}) into account, we obtain
\begin{equation*}
\omega \Big[\beta k^2\int_0^{2\pi}{(\theta''(z))^2dz}+\alpha
\int_0^{2\pi}{(\theta'(z))^2dz}\Big]=2\pi \gamma.
\end{equation*}
By our assumption that $\alpha>0,\beta>0,\gamma>0$ this implies that
$\omega>0$, as stated in Proposition \ref{peri}, and thus we may set
\begin{equation}\label{deflam}
\lambda=\frac{1}{\omega}.
\end{equation}
We also set
\begin{equation}\label{defu}
\theta(z)=z+u(z),
\end{equation}
with $u(z)$ satisfying
\begin{equation}\label{bu}
u(z+2\pi)=u(z)\;\;\; \forall z\in \Real,
\end{equation}
so that we can rewrite (\ref{nu}) as
\begin{equation}
\label{fin} \beta k^2 u'''(z)+\Big(\lambda
k^2-\frac{1}{\lambda}\Big)u''(z)-\alpha u'(z)-\lambda
\sin(z+u(z))+ \lambda \gamma-\alpha=0.
\end{equation}
Rotating fluxon solutions of (\ref{jo}), (\ref{bnd}) thus correspond
to pairs $(\lambda,u(z))$ with $\lambda>0$ and $u(z)$ satisfying
(\ref{bu}),(\ref{fin}).

By integrating (\ref{fin}) over $[0,2\pi]$, taking (\ref{bu}) into
account, we obtain
\begin{equation}\label{int}
\frac{1}{2\pi}\int_0^{2\pi}{\sin(s+u(s))ds}=\gamma-\frac{\alpha}{\lambda},
\end{equation}
so that we can rewrite (\ref{fin}) as
\begin{eqnarray}\label{eq}
\beta k^2 u'''(z)&+&\Big(\lambda
k^2-\frac{1}{\lambda}\Big)u''(z)-\alpha u'(z)\nonumber
\\&-&\lambda \sin(z+u(z))+ \lambda
\frac{1}{2\pi}\int_0^{2\pi}{\sin(s+u(s))ds}=0.
\end{eqnarray}
We note that if $(\lambda,u(z))$ is a solution to our problem then
so is $(\lambda,u(z+c)+c)$ for any constant $c$ (this of course
corresponds to the time-invariance of (\ref{jo})), hence by
adjusting $c$ we may assume that
\begin{equation}\label{intz}
\int_0^{2\pi}{u(s)ds}=0.
\end{equation}
We thus seek pairs $(\lambda,u(z))$, satisfying (\ref{bu}),
(\ref{int}), (\ref{eq}) and (\ref{intz}).

Multiplying (\ref{eq}) by $1+u'(z)$ and integrating over $[0,2\pi]$
we obtain
\begin{lemma}\label{stst}
Any solution $(\lambda,u)$ of (\ref{bu}), (\ref{eq}) and
(\ref{intz}) satisfies
\begin{equation}\label{pos}
\lambda\int_0^{2\pi}{\sin(s+u(s))ds}= \beta k^2
\int_0^{2\pi}{(u''(s))^2ds}+ \alpha \int_0^{2\pi}{(u'(s))^2ds}.
\end{equation}
\end{lemma}
In particular, together with (\ref{int}), (\ref{pos}) implies that
\begin{equation*}
\lambda \gamma\geq \alpha,
\end{equation*}
which gives the upper bound of (\ref{in1}). Also, noting that the
value of the integral on the left-hand side of (\ref{int}) is at
most $1$, we have
\begin{equation}\label{rew}
\gamma-1\leq\frac{\alpha}{\lambda},
\end{equation}
which implies the lower bound in (\ref{in1}). We have thus proven
Proposition \ref{peri}.

\vspace{0.4cm}
We denote by $X,Y,Z$ the Banach spaces of
real-valued functions
$$X=\left\{ u\in H^3[0,2\pi] \; \Big| \; u^{(i)}(0)=u^{(i)}(2\pi),\;\;
0\leq i \leq 2,\;\;\; \int_0^{2\pi}{u(s)ds}=0\right\},$$
$$Y=\left\{ u\in H^1[0,2\pi] \;\Big|\; \int_0^{2\pi}{u(s)ds}=0\right\},$$
$$Z=\left\{ u\in L^2[0,2\pi] \;\Big|\; \int_0^{2\pi}{u(s)ds}=0\right\},$$
with the norms
$$\|u \|_X=\Big( \frac{1}{2\pi}\int_0^{2\pi}{(u'''(s))^2ds} \Big)^{\frac{1}{2}},$$
$$\|u \|_Y=\Big( \frac{1}{2\pi}\int_0^{2\pi}{(u'(s))^2ds} \Big)^{\frac{1}{2}},$$
$$\|u \|_Z=\Big( \frac{1}{2\pi}\int_0^{2\pi}{(u(s))^2ds} \Big)^{\frac{1}{2}}.$$
Assuming $v\in H^1[0,2\pi]$ with $\int_0^{2\pi}{v(s)ds}=0$,
developing $v$ in a Fourier series and using the Parseval identity
one obtains
\begin{equation}\label{wirt}
\int_0^{2\pi}{(v(s))^2ds}\leq \int_0^{2\pi}{(v'(s))^2ds},
\end{equation}
known as Wirtinger's inequality, implying the following inequalities
which we will often use:
$$\| u\|_Y\leq \|u\|_X \;\;\forall u\in
X,\;\;\;\|u\|_Z\leq\|u\|_Y\;\;\forall u\in Y.$$
 We define a family of linear mappings
$L_{\lambda}:X\rightarrow Z$ ($\lambda\neq 0$) by
\begin{equation}\label{defl}
L_{\lambda}(u)=\beta k^2 u'''(z)+ \Big(\lambda
k^2-\frac{1}{\lambda}\Big)u''(z)-\alpha u'(z) \end{equation} and a
nonlinear mapping $N:Z\rightarrow Z$ by
\begin{equation}\label{defn}N(u)=\sin(z+u(z))-
\frac{1}{2\pi}\int_0^{2\pi}{\sin(s+u(s))ds}. \end{equation} Solving
(\ref{bu}), (\ref{eq}), (\ref{intz}) is equivalent to solving
\begin{equation}\label{fa}
L_{\lambda}(u)=\lambda N(u),\;\;\;u\in X,
\end{equation}
so that our problem reduces to: find $(\lambda,u)\in(0,\infty)\times
X$ such that (\ref{fa}) and (\ref{int}) hold (we note that, by a
simple bootstrap argument, any solution $u\in X$ of (\ref{fa}) is in
fact $C^{\infty}$). In other words, defining the nonlinear
functional $\Phi: (0,\infty)\times Z\rightarrow \Real$ by
\begin{equation}
\label{defphi}
\Phi(\lambda,u)=\frac{\alpha}{\lambda}+\frac{1}{2\pi}\int_0^{2\pi}{\sin(s+u(s))ds},
\end{equation}
we need to solve the pair of equations (\ref{fa}) and
\begin{equation}\label{se}
\Phi(\lambda,u)=\gamma,
\end{equation}
for the unknowns $(\lambda,u)\in (0,\infty)\times X.$

\vspace{0.4cm} We derive some estimates on $L_{\lambda}$ which
will be useful. Let $u\in X$ and write $u$ as a Fourier series
$$u(z)=\sum_{l\neq 0}{a_le^{ilz}}.$$
Noting that
$$L_{\lambda}(e^{ilz})=-\Big[(l\alpha+\beta k^2 l^3)i+l^2(\lambda
k^2-\lambda^{-1})\Big]e^{ilz},$$ so that
$$L^{-1}_{\lambda}(e^{ilz})=-\Big[(l\alpha+\beta k^2
l^3)i+l^2(\lambda k^2-\lambda^{-1})\Big]^{-1}e^{ilz},$$ we obtain
\begin{eqnarray*}\| L^{-1}_{\lambda}(u)\|_Y
&=&\| \sum_{l\neq 0}{a_l\Big[(l\alpha+\beta k^2 l^3)i+l^2(\lambda
k^2-\lambda^{-1})\Big]^{-1}e^{ilz}}\|_Y\nonumber\\&=&
\Big(\sum_{l\neq 0}{l^2|a_l|^2\Big[(l\alpha+\beta k^2l^3)^2+l^4
(\lambda
k^2-\lambda^{-1})^2\Big]^{-1}} \Big)^{\frac{1}{2}}\nonumber\\
&=&\Big(\sum_{l\neq 0}{|a_l|^2\Big[(\alpha+\beta
k^2l^2)^2+l^2(\lambda k^2-\lambda^{-1})^2\Big]^{-1}}
\Big)^{\frac{1}{2}}\nonumber\\&\leq& \Big[(\alpha+\beta
k^2)^2+(\lambda
k^2-\lambda^{-1})^2\Big]^{-\frac{1}{2}}\Big[\sum_{l\neq
0}{|a_l|^2}\Big]^{\frac{1}{2}}\nonumber\\&=& \Big[(\alpha+\beta
k^2)^2+(\lambda k^2-\lambda^{-1})^2\Big]^{-\frac{1}{2}}\|u\|_Z.
\end{eqnarray*}
so that
\begin{equation}\label{invb}
\|L_{\lambda}^{-1}\|_{Z,Y}\leq \Big[(\alpha+\beta k^2)^2+(\lambda
k^2-\lambda^{-1})^2\Big]^{-\frac{1}{2}}.
\end{equation}
In particular (dropping the first term on the right-hand side of
(\ref{invb}))
\begin{equation}\label{stt3}
\|L_{\lambda}^{-1}\|_{Z,Y}\leq \frac{\lambda}{|\lambda^2 k^2-1|}.
\end{equation}

We can also obtain
\begin{eqnarray*}\| L^{-1}_{\lambda}(u)\|_X&=&\Big(\sum_{l\neq
0}{l^6|a_l|^2\Big[(l\alpha+\beta k^2l^3)^2+l^4(\lambda
k^2-\lambda^{-1})^2\Big]^{-1}} \Big)^{\frac{1}{2}}\nonumber\\&\leq&
\frac{\lambda}{|\lambda^2 k^2-1|} \Big(\sum_{l\neq 0}{l^2|a_l|^2}
\Big)^{\frac{1}{2}}= \frac{\lambda}{|\lambda^2 k^2-1|}\|u\|_Y,
\end{eqnarray*}
so that
\begin{equation}\label{invb2}
\|L_{\lambda}^{-1}\|_{Y,X}\leq \frac{\lambda}{|\lambda^2 k^2-1|}.
\end{equation}

Defining $F:(0,\infty)\times Z\rightarrow Z$ by
\begin{equation}\label{deff} F(\lambda,u)=\lambda
L_{\lambda}^{-1}\circ N(u),
\end{equation}
we may rewrite (\ref{fa}) as
\begin{equation}\label{fixp}
u=F(\lambda,u),\;\;\;u\in Z.
\end{equation}
We note that since $L_{\lambda}^{-1}$ maps $Z$ onto $X$, any
solution $u\in Z$ of (\ref{fixp}) is in fact in $X$.

Since $\|\sin(z+u(z))\|_{L^2}\leq1$ and $N(u)$ is the orthogonal
projection of $\sin(z+u(z))\in L^2[0,2\pi]$ into $Z$, we have
\begin{equation}\label{bndn}
\| N(u) \|_Z\leq 1\;\;\;\forall u\in Z.
\end{equation}
Using (\ref{bndn}) and (\ref{invb}), we have the bound:
\begin{equation}\label{st}
\|F(\lambda,u)\|_Z\leq \|F(\lambda,u)\|_Y\leq \lambda
\Big[(\alpha+\beta k^2)^2+(\lambda
k^2-\lambda^{-1})^2\Big]^{-\frac{1}{2}}\;\;\;\forall u\in Z,\;
\lambda>0.
\end{equation}

We extend the definition of $F(\lambda,u)$ by setting $F(0,u)=0$
for all $u\in Z$ (this is necessary because $L_{\lambda}$ is
undefined for $\lambda=0$). The bound (\ref{st}) shows that the
extension is continuous, and moreover that the mapping $F$ takes
bounded sets in $[0,\infty)\times Z$ into bounded sets in $Y$,
hence (since the embedding of $Y$ into $Z$ is compact) into
compact sets in $Z$, so that $F:[0,\infty)\times Z\rightarrow Z$
is a compact mapping.

We define
$$\Sigma =\{ (\lambda,u)\in
(0,\infty)\times Z\;|\; u=F(\lambda,u)\}.$$
 The solutions of
(\ref{fa}),(\ref{se}), are the solutions of the equation
\begin{equation}\label{red}
\Phi(\lambda,u)=\gamma, \;\;\;(\lambda,u)\in \Sigma.
\end{equation}
In other terms, recalling (\ref{deflam}), the bias-frequency
characteristic $\chi$ defined in the introduction is given by
\begin{equation}\label{exchi}
\chi=\{(\omega,\Phi(\omega^{-1},u))\;|\;(\omega^{-1},u)\in \Sigma\;
\}.
\end{equation}

We will now apply the following version of the Leray-Schauder
continuation principle (see, {\it{e.g.}}, \cite{zeidler}, Theorem
14.D):

\begin{theorem}\label{top}
Let $Z$ be a Banach space and let $F:[0,\infty)\times Z\rightarrow
Z$ be a continuous and compact mapping, with
$$F(0,u)=0\;\;\; \forall u\in Z.$$
Define $\Sigma\subset (0,\infty)\times Z$ by
 $$\Sigma =\{ (\lambda,u)\in
(0,\infty)\times Z\;|\; u=F(\lambda,u)\}.$$ Then there exists an
unbounded connected set $\Sigma_0\subset \Sigma$ with $(0,0)\in
\overline{\Sigma}_0$.
\end{theorem}
Since the mapping $F$ that we have defined satisfies the hypotheses
of Theorem \ref{top}, we have a set $\Sigma_0$ as in the Theorem.

We define $\chi_0\subset \chi$ by
\begin{equation}\label{defchi0}
\chi_0=\{(\omega,\Phi(\omega^{-1},u))\;|\;
\omega>0,\;(\omega^{-1},u)\in \Sigma_0\},
\end{equation}
and we will show that $\chi_0$ has all the properties claimed in
Theorem \ref{curt}.

Since $\Phi$ is continuous and $\Sigma_0$ is connected, $\chi_0$ is
connected, so we have (i) of Theorem \ref{curt}.

Part (ii) of Theorem \ref{curt} follows from the following
\begin{lemma}\label{fal}
Let $\Sigma_0$ be as in Theorem \ref{top}. For any $\lambda>0$ there
exists $u\in Z$ such that $(\lambda,u)\in \Sigma_0$.
\end{lemma}

\noindent {\sc{proof:}} Since from (\ref{fixp}) and (\ref{st}) we
have
\begin{equation}\label{bbb}(\lambda,u)\in \Sigma\Rightarrow \|u\|_Z\leq
\lambda \frac{1}{\beta k^2+\alpha},\end{equation} we conclude that
the unboundedness of $\Sigma_0$ is in the $\lambda$-direction, that
is, for any $M>0$ there exists $(\lambda,u)\in \Sigma_0$ with
$\lambda>M$. In other words, denoting by $\Lambda\subset (0,\infty)$
the projection of $\Sigma_0$ to the $\lambda$-axis,
$$\Lambda=\{ \lambda>0\;|\; \exists u\in Z {\mbox{ such that }}(\lambda,u)\in \Sigma_0\},$$
 we have that $\Lambda$
contains arbitrarily large values. Since $(0,0)\in
\overline{\Sigma}_0$, $\Lambda$ also contains arbitrary small
positive values. Since $\Sigma_0$, hence $\Lambda$, is connected, we
conclude that $\Lambda=(0,\infty)$. $\square$

To prove that $\chi_0$ satisfies property (iii) of Theorem
\ref{curt} (and hence also Theorem \ref{main}), we need to show that
\begin{equation}\label{cont}
(0,\infty) \subset \Phi(\Sigma_0).
\end{equation}
Since $\Sigma_0$ is connected and $\Phi$ is continuous,
$\Phi(\Sigma_0)$ is also connected, so that to prove (\ref{cont})
it suffices to show that $\Phi$ takes arbitrarily large and
arbitrarily small positive values on $\Sigma_0$. Since the
integral in the definition (\ref{defphi}) of $\Phi$ is bounded by
$1$, it is immediate that
\begin{equation}\label{lim0}
\lim_{\lambda\rightarrow 0+,\; (\lambda,u)\in
\Sigma}{\Phi(\lambda,u)}=+\infty, \end{equation} so that $\Phi$
takes arbitrarily large values on $\Sigma_0$. We shall prove
\begin{lemma}\label{linf}
\begin{equation}\label{lim}
\lim_{\lambda\rightarrow +\infty,\; (\lambda,u)\in
\Sigma}{\Phi(\lambda,u)}=0.
\end{equation}
\end{lemma}
This implies that $\Phi$ takes arbitrarily small positive values on
$\Sigma_0$, completing the proof of (\ref{cont}), and thus of part
(iii) of Theorem \ref{curt} holds.

\vspace{0.4cm}

\noindent {\sc{proof of Lemma \ref{linf}:}} Applying Wirtinger's
inequality (\ref{wirt}) with $v=u'$, Lemma \ref{stst} implies
\begin{equation}\label{nn1}
(\lambda,u)\in \Sigma \;\Rightarrow \;\lambda\frac{1}{2\pi}\Big|
\int_0^{2\pi}{\sin(s+u(s))ds}\Big|\leq (\beta k^2 +\alpha)
\frac{1}{2\pi}\int_0^{2\pi}{(u'(s))^2ds}.
\end{equation}
From (\ref{fixp}) and (\ref{stt3}) we obtain
\begin{equation}\label{nn2} (\lambda,u)\in\Sigma \;\Rightarrow\;
\|u\|_Y\leq \frac{\lambda^2}{|\lambda^2k^2-1|}.\end{equation}
(\ref{nn1}) together with (\ref{nn2}) implies
$$(\lambda,u)\in\Sigma \;\Rightarrow\; \Big| \frac{1}{2\pi}\int_0^{2\pi}{\sin(s+u(s))ds}\Big| \leq
\frac{(\beta k^2 +\alpha)\lambda^3}{(\lambda^2k^2-1)^2},$$ which
implies that
\begin{equation}\label{plus}
\lim_{\lambda \rightarrow +\infty,\;
(\lambda,u)\in
\Sigma}{\frac{1}{2\pi}\int_0^{2\pi}{\sin(s+u(s))ds}}=0,
\end{equation}
hence we have (\ref{lim}). $\square$

\vspace{0.4cm} To prove claim (iv) of Theorem \ref{curt}, let
$\omega_n=\frac{1}{n}$ ($n\geq 1$), and note that by part (ii) of
the Theorem, there exists, for each $n$,  $\gamma_n>0$ such that
$(\omega_n,\gamma_n)\in \chi_0$. By the definition of $\chi_0$ there
exists $u_n\in \Sigma_0$ such that
$\gamma_n=\Phi(\omega_n^{-1},u_n)$. But by Lemma \ref{linf} we have
$\gamma_n\rightarrow 0$ as $n\rightarrow \infty$. Hence we have
$(\omega_n,\gamma_n)\rightarrow (0,0)$, so we have $(0,0)\in
\overline{\chi_0}$.

To prove part (v) of Theorem \ref{curt}, we will use the following
two lemmas, which will also be used in the following sections.

\begin{lemma}\label{sl1} If
\begin{equation}\label{contract}\lambda \Big[(\alpha+\beta k^2)^2+(\lambda
k^2-\lambda^{-1})^2\Big]^{-\frac{1}{2}}<1
\end{equation}
then $F(\lambda,.):Z\rightarrow Y$ (hence also
$F(\lambda,.):Z\rightarrow Z$) is a contraction.
\end{lemma}

\noindent {\sc{proof:}} Computing the Frech\'et derivative of
$N:Z\rightarrow Z$ we have
$$N'(u)(v)=\cos(z+u(z))v(z)-
\frac{1}{2\pi}\int_0^{2\pi}{\cos(s+u(s))v(s)ds},$$ so that $N'(u)$
is the mapping from $Z$ to $L^2[0,2\pi]$ given by $v(z)\rightarrow
\cos(z+u(z))v(z)$ followed by an orthogonal projection into the
subspace $Z$, and this implies
\begin{equation}\label{cont1}
\|N'(u)\|_{Z,Z}\leq 1 \;\;\;\forall u\in Z.
\end{equation}
From (\ref{invb}), (\ref{deff}) and (\ref{cont1}) we conclude that
\begin{equation}\label{fdb}\|D_uF(\lambda,u)\|_{Z,Y}=\lambda\| L_{\lambda}^{-1}N'(u)\|_{Z,Y}\leq \lambda \Big[(\alpha+\beta
k^2)^2+(\lambda k^2-\lambda^{-1})^2\Big]^{-\frac{1}{2}}\;\;\;\forall
u\in Z,
\end{equation}
which, by (\ref{contract}), completes the proof. $\square$

\begin{lemma}
\label{sl} Fixing $\alpha\geq 0$, $\beta> 0$, there exists
$\lambda_0>0$ such that for any $\lambda\in [0,\lambda_0)$ the
equation (\ref{fixp}) has a {\bf{unique}} solution, which we
denote by $u_{\lambda}$. The mapping $\lambda\rightarrow
u_{\lambda}$ from $[0,\lambda_0)$ into $Z$ is continuous and is
real-analytic for $\lambda\in (0,\lambda_0)$.
\end{lemma}

\noindent {\sc{proof:}}  It is easy to see that we can find
$\lambda_0>0$ such that (\ref{contract}) holds for $\lambda\in
[0,\lambda_0)$, and thus by Lemma \ref{sl1}, for any $\lambda\in
[0,\lambda_0)$ the mapping $F(\lambda,.):Z\rightarrow Z$ is a
contraction, so that, by Banach's fixed-point Theorem, equation
(\ref{fixp}) has a {\bf{unique}} solution, which we denote by
$u_{\lambda}$. Real-analyticity of the mapping $\lambda\rightarrow
u_{\lambda}$ follows from the real-analytic implicit function
Theorem. $\square$

\vspace{0.4cm} Returning to the proof of part (v) of Theorem
\ref{curt}, we note that since, by Lemma \ref{fal}, for each
$\lambda$ there exists $u$ such that $(\lambda,u)\in \Sigma_0$, and
since  the uniqueness in Lemma \ref{sl} implies that when
$\lambda<\lambda_0$ such a $u$ is unique in $\Sigma$, we must have
\begin{equation}\label{imunb}
0<\lambda<\lambda_0,\; (\lambda,u)\in\Sigma \;\Rightarrow\;
(\lambda,u)\in \Sigma_0.
\end{equation}
Thus if we set $\omega_0=\frac{1}{\lambda_0}$ we have by
(\ref{imunb}), (\ref{exchi}), (\ref{defchi0})
\begin{equation}\label{cons}
\omega>\omega_0,\; (\omega,\gamma)\in \chi \;\Rightarrow\;
(\omega,\gamma)\in \chi_0.
\end{equation}
Setting $\gamma_0=\alpha\omega_0+1$ and using Proposition \ref{peri}
and (\ref{cons}) we obtain
\begin{equation}\label{cons1}
\gamma>\gamma_0,\; (\omega,\gamma)\in \chi \;\Rightarrow\;
\omega>\omega_0,\; (\omega,\gamma)\in \chi \;
\Rightarrow\;(\omega,\gamma)\in \chi_0,
\end{equation}
and (\ref{cons}),(\ref{cons1}) mean that
$$(\omega,\gamma)\in \chi\setminus\chi_0\;\Rightarrow\; \omega \leq \omega_0, \; \gamma\leq \gamma_0,$$
so that $\chi\setminus \chi_0$ is bounded.

\vspace{0.4cm} We note for later use that (\ref{pos}) implies that
\begin{lemma} \label{posi}
$$(\lambda,u)\in \Sigma,\;\; \lambda>0 \Rightarrow
\frac{1}{2\pi}\int_0^{2\pi}{\sin(s+u(s))ds}>0.$$\end{lemma} The
strictness in the above inequality follows from the fact that, by
(\ref{pos}), if we had equality this would imply $u=0$, but it is
easy to check that $(\lambda,0)\not\in\Sigma$ for $\lambda>0$.

\section{More on the bias-frequency characteristic}
\label{smoothness}

In this section we prove Theorem \ref{connect}, which shows that for
a certain open subset of the space of parameters $L,\alpha,\beta$,
the bias-frequency characteristic $\chi$ is a connected, real
analytic curve, and can be represented as the graph of a function:
$\gamma=\overline{\gamma}(\omega)$.

\begin{lemma}\label{opo}
Assume that $\alpha\geq 0$, $\beta>0$ and the inequalities
(\ref{k1}) and (\ref{incon2}) hold. Then for all $\lambda\geq 0$,
the equation (\ref{fixp}) has a {\bf{unique}} solution, which we
denote by $u_{\lambda}$, and therefore
\begin{equation}\label{sec}
\Sigma=\Sigma_0=\{ (\lambda,u_{\lambda})\;|\; \lambda\geq 0\}.
\end{equation}
The mapping $\lambda\rightarrow u_{\lambda}$ from $[0,\infty)$
into $Y$ is continuous, and real-analytic on $(0,\infty)$.
\end{lemma}

\noindent {\sc{proof:}} We use Banach's fixed point Theorem. By
 Lemma \ref{sl1}, to show that $F(\lambda,.)$ is a contraction from $Z$ to
 itself for
all $\lambda\geq 0$ it suffices to show that (\ref{contract})
holds for all $\lambda\geq 0$. After some manipulations, and
setting $\mu=\lambda^2$, we see that this inequality is equivalent
to
\begin{equation}\label{inw}
p(\mu)\equiv (k^4-1)\mu^2+[(\alpha+\beta
k^2)^2-2k^2]\mu+1>0\;\;\;\forall \mu\geq 0.\end{equation} The
condition $k>1$ ensures that the quadratic function $p(\mu)$ has a
minimum, and we have $p(0)=1>0$. Moreover the condition
(\ref{incon2}) ensures that $p'(0)\geq 0$, which implies $p$ is
increasing on $[0,\infty)$, so that (\ref{inw}) holds.  We thus
obtain a unique solution $u_{\lambda}$ to (\ref{fixp}). The fact
that it is real-analytic in $\lambda\in (0,\infty)$ follows from the
analytic implicit function Theorem and the fact that the function
$\lambda\rightarrow L_{\lambda}^{-1}$ from $(0,\infty)$ to the space
$B(Z)$ of bounded linear operators from $Z$ to itself is
real-analytic. $\square$

\vspace{0.4cm} Let us note here that it is precisely the last
statement in the above proof - the fact that $\lambda\rightarrow
L_{\lambda}^{-1}$ is real-analytic, that breaks down when $\beta=0$,
and is responsible for the fact that in that case ({\it{i.e.}} the
case of the damped, dc-forced sine-Gordon equation) the curve $\chi$
is {\it{not}} smooth (see \cite{watanabe}, fig. 2). The reason that,
when $\beta=0$, $\lambda\rightarrow L_{\lambda}^{-1}$ is not
real-analytic can be gleaned by recalling the definition
(\ref{defl}): if $\beta=0$ then at $\lambda=\frac{1}{k}$ the
coefficient of the highest derivative in the operator $L_{\lambda}$
vanishes.

\vspace{0.4cm} Thus, under the assumptions of Lemma \ref{opo}, we
have
$$\chi=\chi_0=\{(\omega,\Psi(\omega^{-1}))\;|\;\omega>
0\},$$ where $\Psi:(0,\infty)\rightarrow\Real$ is defined by
\begin{equation}\label{defpsi}\Psi(\lambda)=\Phi(\lambda,u_{\lambda}),
\end{equation}
so defining $\overline{\gamma}(\omega)=\Psi(\omega^{-1})$ we have
(\ref{rc}). The fact that $\overline{\gamma}(\omega)>0$ for
$\omega>0$ follows from Lemma \ref{posi}. (\ref{st1}) follows from
Lemma \ref{linf}, and (\ref{st2}) follows from (\ref{lim0}). This
concludes the proof of Theorem \ref{connect}.

\section{uniqueness of rotating fluxon waves for large bias current}
\label{uniqueness}

We now prove Theorem \ref{uni1}. By Lemma \ref{sl}, defining
$\lambda_0$ as in Lemma \ref{sl}, for any $\lambda\in [0,\lambda_0)$
the equation (\ref{fixp}) has a {\bf{unique}} solution, which we
denote by $u_{\lambda}$. We define the function
$\Psi:(0,\lambda_0)\rightarrow \Real$ by (\ref{defpsi}), where
$\Phi$ is the functional defined by (\ref{defphi}), so that
solutions of (\ref{red}) with $\lambda\in (0,\lambda_0)$ correspond
to solutions of
\begin{equation}\label{red1}
\Psi(\lambda)=\gamma.
\end{equation}
We show below that there exists some $\lambda_1\in (0,\lambda_0)$
such that
\begin{equation}\label{dec}
\lambda\in (0,\lambda_1) \;\;\Rightarrow\;\;\Psi'(\lambda)<0.
\end{equation}
We note that if $\gamma$ is sufficiently large then, by (\ref{in1}),
we have $\lambda\in (0,\lambda_1)$ for any solution of (\ref{red1}).
Therefore (\ref{dec}) implies uniqueness of solutions of
(\ref{red1}), hence uniqueness of the rotating fluxon wave, for
sufficiently large $\gamma$, which is the content of Theorem
\ref{uni1}.

To prove claim (\ref{dec}) we note that, since
$$\Psi(\lambda)=\frac{\alpha}{\lambda}+\frac{1}{2\pi}\int_0^{2\pi}{\sin(s+u_{\lambda}(s))ds},
$$
we have $$\Psi'(\lambda)=-\frac{\alpha}{\lambda^2}
+\frac{1}{2\pi}\int_0^{2\pi}{\cos(s+u_{\lambda}(s))D_{\lambda}u_{\lambda}(s)ds},$$
and using Cauchy's inequality
\begin{equation} \label{dps}
\Psi'(\lambda)\leq
-\frac{\alpha}{\lambda^2}+\|D_{\lambda}u_{\lambda}\|_Z,
\end{equation}
so to prove (\ref{dec}) it suffices to show that
$\|D_{\lambda}u_{\lambda}\|_Z=o(\frac{1}{\lambda^2})$ as
$\lambda\rightarrow 0+$. In fact we shall prove a much stronger
statement:

\begin{lemma}\label{tec}
We have
\begin{equation}\label{bnda2}\|D_{\lambda}u_{\lambda}\|_Z=O(\lambda)\;\;\;as\;\lambda\rightarrow
0+.
\end{equation}
\end{lemma}

\noindent {\sc{proof:}} We differentiate the identity
$$L_{\lambda}(u_{\lambda})=\lambda N(u_{\lambda})$$
with respect to $\lambda$, obtaining
\begin{equation}\label{difl}(D_{\lambda}L_{\lambda})(u_{\lambda})+L_{\lambda}(D_{\lambda}u_{\lambda})
=N(u_{\lambda})+\lambda
N'(u_{\lambda})(D_{\lambda}u_{\lambda}).\end{equation} We note that,
because we assume $\lambda\in(0,\lambda_0)$, where $\lambda_0$ is
defined as in Lemma \ref{sl}, $\lambda
L_{\lambda}^{-1}N'(u_{\lambda})$ is a contraction from $Z$ to itself
so the inverse of $I-\lambda L_{\lambda}^{-1}N'(u_{\lambda})$ is
well defined as a mapping from $Z$ to itself. Moreover we have,
using (\ref{fdb}),
\begin{equation}\label{es1}
\|[I-\lambda L_{\lambda}^{-1}N'(u_{\lambda})]^{-1}\|_{Z,Z} \leq
\frac{1}{1-\lambda \|L_{\lambda}^{-1}N'(u_{\lambda})\|_{Z,Z}}=O(1)
\;\;\;as\;\lambda\rightarrow 0+.
\end{equation}
We can rewrite (\ref{difl}) in the form
\begin{equation*}\label{dul}
D_{\lambda}u_{\lambda}=[I-\lambda
L_{\lambda}^{-1}N'(u_{\lambda})]^{-1}L_{\lambda}^{-1}[N(u_{\lambda})
-(D_{\lambda}L_{\lambda})(u_{\lambda})],
\end{equation*}
which implies
\begin{eqnarray}
\label{bnda1} &&\|D_{\lambda}u_{\lambda}\|_Z \\  &\leq&
\|[I-\lambda
L_{\lambda}^{-1}N'(u_{\lambda})]^{-1}\|_{Z,Z}\Big[\|L_{\lambda}^{-1}N(u_{\lambda})\|_{Z}+
\|L_{\lambda}^{-1}(D_{\lambda}L_{\lambda})(u_{\lambda})\|_Z\Big].\nonumber
\end{eqnarray}

Using (\ref{nn2}) we obtain
\begin{eqnarray*}
&&\|N(u_{\lambda})\|_Y=\Big(\frac{1}{2\pi}\int_0^{2\pi}{\cos^2(s+u_{\lambda}(s))(1+u_{\lambda}'(s))^2ds}\Big)^{\frac{1}{2}}
\nonumber\\
&\leq&\Big(\frac{1}{2\pi}\int_0^{2\pi}{(1+u_{\lambda}'(s))^2ds}\Big)^{\frac{1}{2}}=[1+\|u_{\lambda}\|_Y^2]^{\frac{1}{2}}
\leq \Big[1+\Big(
\frac{\lambda^2}{\lambda^2k^2-1}\Big)^2\Big]^{\frac{1}{2}},
\end{eqnarray*}
so that
\begin{equation}\label{sof}
\|N(u_{\lambda})\|_Y=O(1)\;\;\;as\;\lambda\rightarrow 0+.
\end{equation}
From (\ref{invb2}) and (\ref{sof}) we have
\begin{equation}\label{b1}
\|L_{\lambda}^{-1}N(u_{\lambda})\|_{Z}\leq
\|L_{\lambda}^{-1}N(u_{\lambda})\|_{X}\leq
\|L_{\lambda}^{-1}\|_{Y,X}\|N(u_{\lambda})\|_Y=O(\lambda)\;\;\;as\;\lambda\rightarrow
0+.
\end{equation}

From the definition (\ref{defl}) of $L_{\lambda}$ we have, for all
$u\in X$
$$
(D_{\lambda}L_{\lambda})(u)=\Big(k^2+\frac{1}{\lambda^2}\Big)u''(z),
$$
hence
\begin{equation}\label{ld}
\|(D_{\lambda}L_{\lambda})(u)\|_Y
=\Big(k^2+\frac{1}{\lambda^2}\Big)\|u\|_X\;\;\;\forall u\in X.
\end{equation}
From (\ref{b1}) we have
\begin{equation}\label{bn2}
\|u_{\lambda}\|_X=\lambda\|L_{\lambda}^{-1}N(u_{\lambda})\|_X =
O(\lambda^2)\;\;\;as\;\lambda\rightarrow 0+,
\end{equation}
so using (\ref{ld}),(\ref{bn2})
\begin{equation}\label{ld1}
\|(D_{\lambda}L_{\lambda})(u_{\lambda})\|_Y =
O(1)\;\;\;as\;\lambda\rightarrow 0+,
\end{equation}
and together with (\ref{invb2})
\begin{equation}\label{ld2}
\|L_{\lambda}^{-1}(D_{\lambda}L_{\lambda})(u_{\lambda})\|_Z\leq
\|L_{\lambda}^{-1}(D_{\lambda}L_{\lambda})(u_{\lambda})\|_X=O(\lambda)\;\;\;as\;\lambda\rightarrow
0+.
\end{equation}
Using (\ref{es1}), (\ref{bnda1}), (\ref{b1}) and (\ref{ld2}) we
obtain (\ref{bnda2}). $\square$

\section{multiplicity of rotating fluxon waves}
\label{multiplicity}

In this section we prove Theorem \ref{multi}, which shows that for
some values of the parameters $L$,$\beta$, if $\alpha>0$ is
sufficiently small then there exists an interval of $\gamma$'s for
which there exist at least three rotating fluxon waves.

\begin{lemma}\label{opo1}
Assume that $\beta>0$ and (\ref{k1}),(\ref{lcond2}) hold. Then for
any $\alpha\geq 0$, $\lambda\geq 0$, the equation (\ref{fixp}) has a
{\bf{unique}} solution, which we denote by $u_{\alpha,\lambda}$. The
mapping $(\alpha,\lambda)\rightarrow u_{\alpha,\lambda}$ from
$[0,\infty)\times [0,\infty)$ into $Z$ is continuous.
\end{lemma}

As in the proof of Lemma \ref{opo}, this follows from the Banach
fixed point Theorem applied to (\ref{fixp}). Note that
(\ref{lcond2}) implies that (\ref{incon2}) holds for all $\alpha\geq
0$.

Since in the following argument the dependence on $\alpha$ is
important, we display it in the notation (of course
$u_{\alpha,\lambda}$ also depends on the parameters $k$ and
$\beta$, but since we shall regard these as fixed we can suppress
this dependence).

\vspace{0.4cm} \noindent {\sc{proof of Theorem \ref{multi}:}} Under
the conditions of Lemma \ref{opo1} we can define
$\Psi_{\alpha}(\lambda)$ as in (\ref{defpsi}), now defined for all
$\lambda\in [0,\infty)$, so that rotating fluxon waves correspond to
solutions of
\begin{equation}\label{eqs}
\Psi_{\alpha}(\lambda)=\gamma, \;\;\;\lambda>0.
\end{equation}

We write $\Psi_{\alpha}(\lambda)$ in the form (see (\ref{defphi}))
\begin{equation}\label{psif}
\Psi_{\alpha}(\lambda)=\frac{\alpha}{\lambda}+\Upsilon_{\alpha}(\lambda),
\end{equation}
where
$$\Upsilon_{\alpha}(\lambda)=\frac{1}{2\pi}\int_0^{2\pi}{\sin(s+u_{\alpha,\lambda}(s))ds}.$$

We note some properties of the function
$\Upsilon_{\alpha}(\lambda)$ (valid for any $\alpha\geq 0$). Since
$u_{\alpha,0}=0$ we have
\begin{equation}\label{lz}
\Upsilon_{\alpha}(0)=0.
\end{equation}
From (\ref{plus}) we obtain:
\begin{equation}\label{li}
\lim_{\lambda\rightarrow +\infty} \Upsilon_{\alpha}(\lambda)=0.
\end{equation}
Lemma \ref{posi} implies
\begin{equation}\label{ppos}
\Upsilon_{\alpha}(\lambda)>0 \;\;\;\forall \lambda>0.
\end{equation}
(\ref{lz}),(\ref{li}),(\ref{ppos}), imply that
$\Upsilon_{\alpha}(\lambda)$ attains a positive global maximum at
some $\lambda>0$. In particular this is true for $\alpha=0$, and
we denote the global maximum of $\Upsilon_0(\lambda)$ by
$\gamma_0$, and assume that it is attained at $\lambda_0>0$:
\begin{equation}\label{max}
\Upsilon_0(\lambda_0)=\gamma_0=\max_{\lambda>0}{\Upsilon_0(\lambda)}.
\end{equation}
We now fix an arbitrary $0<\epsilon<\frac{1}{2}\gamma_0$. By
(\ref{lz}),(\ref{li}) we can choose $0<\lambda_1<\lambda_0$ and
$\lambda_2>\lambda_0$ so that
\begin{equation}\label{l1}
\Upsilon_0(\lambda_1)<\frac{\epsilon}{4},\;\;\;
\Upsilon_0(\lambda_2)<\frac{\epsilon}{4}.
\end{equation}
Since the dependence of $\Upsilon_{\alpha}(\lambda)$ on $\alpha$
is continuous, we can choose $\alpha_0>0$ sufficiently small so
that
\begin{equation}\label{asm}
|\Upsilon_{\alpha}(\lambda)-\Upsilon_{0}(\lambda)|<\frac{\epsilon}{2}\;\;\;\forall
\alpha\in [0,\alpha_0],\lambda\in [0,\lambda_2].
\end{equation}
From (\ref{max}),(\ref{l1}) and (\ref{asm}) we obtain, for all
$\alpha\in [0,\alpha_0]$,
\begin{equation}\label{m1}
\Upsilon_{\alpha}(\lambda_0)>\gamma_0-\frac{\epsilon}{2},
\end{equation}
\begin{equation}\label{m2}
\Upsilon_{\alpha}(\lambda_1)<\frac{3}{4}\epsilon,\;\;\;
\Upsilon_{\alpha}(\lambda_2)<\frac{3}{4}\epsilon.
\end{equation}
Moreover we can choose $\alpha_0>0$ sufficiently small so that
\begin{equation}\label{ers}
\frac{\alpha_0}{\lambda_1}<\frac{\epsilon}{2}.
\end{equation}
From (\ref{psif}) and (\ref{ers}) we then obtain
$$|\Psi_{\alpha}(\lambda)-\Upsilon_{\alpha}(\lambda)|<\frac{\epsilon}{2}\;\;\;\forall
\alpha\in [0,\alpha_0],\lambda\geq \lambda_1.$$ Together with
(\ref{m1}),(\ref{m2}) this gives, for all $0<\alpha<\alpha_0$,
\begin{equation}\label{mm1}
\Psi_{\alpha}(\lambda_0)>\gamma_0-\epsilon,
\end{equation}
\begin{equation}\label{mm2}
\Psi_{\alpha}(\lambda_1)<\epsilon,
\end{equation}
\begin{equation}\label{mm3}
\Psi_{\alpha}(\lambda_2)<\epsilon.
\end{equation}
and by (\ref{psif}) we also have
\begin{equation}\label{mm4}
\lim_{\lambda\rightarrow 0+}{\Psi_{\alpha}(\lambda)}=+\infty.
\end{equation}
Thus by continuity of $\Psi_{\alpha}$ on $(0,\infty)$, if we take
$\gamma\in (\epsilon,\gamma_0-\epsilon)$ then by (\ref{mm4}) and
(\ref{mm2}) the equation (\ref{eqs}) has a solution $\lambda\in
(0,\lambda_1)$, by (\ref{mm2}) and (\ref{mm1}) the equation
(\ref{eqs}) has a solution $\lambda\in (\lambda_1,\lambda_0)$, and
by (\ref{mm1}) and (\ref{mm3}) the equation (\ref{eqs}) has a
solution $\lambda\in (\lambda_1,\lambda_2)$. We have thus found,
under the stated assumptions, three solutions of (\ref{eqs}),
corresponding to three rotating fluxon waves with different
frequencies. $\square$

\end{document}